\documentclass[preprint,aps,12pt,showpacs,nofootinbib,tightenlines]{revtex4}
\usepackage{mathrsfs}
\usepackage{amsmath}
\usepackage{amssymb}
\usepackage{epsfig}
\usepackage{graphicx}
\newcommand{\psl}{ P \hspace{-2.4truemm}/ }

\newcommand{\esl}{ \epsilon \hspace{-1.5truemm}/ }

\def\be{\begin{eqnarray}}
\def\en{\end{eqnarray}}
\def\non{\nonumber\\}

\def\prd{{Phys. Rev. D}~}
\def\prl{{ Phys. Rev. Lett.}~}
\def\plb{{ Phys. Lett. B}~}
\def\npb{{ Nucl. Phys. B}~}
\def\epjc{{ Eur. Phys. J. C}~}

\begin{document}
\title{Three body radiative decay $B_s\to \phi \bar K^0 \gamma$ in the PQCD approach}
\author{Zhi-Qing Zhang$^1$, Hongxia Guo\footnote{Corresponding author: Hongxia Guo, e-mail: guohongxia@zzu.edu.cn.}$^2$}
\affiliation{\it \small $^1$  Department of Physics, Henan
University of
Technology,\\ \small \it Zhengzhou, Henan 450052, P. R. China; \\
\small $^2$ \it School of Mathematics and Statistics, Zhengzhou University,\\
\small \it Zhengzhou, Henan 450001, P. R. China } 
\date{\today}
\begin{abstract}
We study the three body radiative decay $B_s\to \phi \bar K^0 \gamma$ by introducing the $\phi K$ pair distribution amplitudes (DAs) in the perturbative QCD approach. This nonperturbative
inputs, the two meson DAs, is very important to simplify the calculations. Besides the dominant electromagnetic penguin operator $O_{7\gamma}$, the subleading contributions from chromomagnetic
penguin operator $O_{8g}$, quark-loop corrections and annihilation type amplitudes are also considered. We find that the branching ratio for the decay $B_s\to \phi \bar K^0 \gamma$ is
about $(9.26^{+1.79+3.12+0.64}_{-1.61-3.86-0.49})\times10^{-8}$, which is much smaller compared with that for the decay $B^0\to \phi K^0\gamma$. It is mainly because that the former decay induces by
$b\to d\gamma$ with small CKM matrix element being proportional to $\lambda^3$.
The prediction for the direct CP asymmetry is $A^{dir}_{CP}(B_s\to \phi \bar K^0 \gamma)=(-4.1^{+0.4+1.7+0.2}_{-0.6-1.2-0.1})\%$, which is well consistent with the result from the U-spin symmetry
approach. we also predict the $B_s \to\phi \bar K^0\gamma$ decay spectrum, which exhibits a maximu at the $\phi K$ invariant masss around 1.95 GeV.

\end{abstract}

\pacs{13.25.Hw, 12.38.Bx, 14.40.Nd}
\vspace{1cm}

\maketitle


\section{Introduction}\label{intro}
In the past decades, the $B$ meson three body decays have attracted lots of attentions both on experiments and theories. On the experimental side, many data for the three-body $B$ meson decays have been
measured by different collaborations, such as Belle \cite{belle0,belle1,belle2,belle3,belle4}, BaBar \cite{babar0,babar1,babar2,babar3}, which provide valuable information about complicated strong dynamics. Especially large
regional CP violation in $KKK,K\pi\pi,KK\pi,\pi\pi\pi$ final states \cite{lhc1,lhc2,lhc3} are revealed. All of these have raised great interests and severe challenge to the theorists.
On the theoretical side, substantial progress on three-body $B$ meson decays has been made through the symmetry principles and factorization theorems. The former includes Isospin, U-spin, flavor
SU(3) symmetries and CPT invariance \cite{gronau1,gronau2,dxu1,xghe,imb}. The later includes the naive factorization \cite{wshou1,wshou2,wshou3}, QCD factorization (QCDF) \cite{furman,eibennich,leitner,cheng1,cheng2,cheng3,yingli,cheng4}
and the perturbative QCD approach \cite{lihn0,lihn1,lihn2,yli,ajma,zrui,zrui1,hnli3,yli2,ajma2,yli3,ajma3,yli4,ajma4,hnli4,cwang}. It is noticed that the rigorous justification of
these approach for the three body decays is not yet available, so these approaches worked in the phenomentological factorization. Compared with two-body decays, the three-body decays is much more
complicated because of receiving  both nonresonant and resonant contributions. It is difficult to separated them clearly \cite{krankl}. Furthermore, the final state interactions (FSIs) may be more
significant in the three-body decays than those in the two-body decays. Because there exist two distinct FSIs mechanisms. One is the interactions between the meson pair in the resonant region
associated with various intermediate states. The other is the rescattering between the third particle (usually referred to as "bachelor") and the pair of mesons. The most difficult for the three body
decays is evaluation of the matrix elements for B meson transition into two hadrons.
Enormous number of diagrams will need to be calculated if one evaluates directly the hard kernels for the three-body decays, which contain two virtual gluons at lowest order.
Fortunately, the region with the two gluons being hard simultaneously is power suppressed and not important, which corresponds to the central region of a Dalitz plot \cite{cheng4}.
The dominant contributions come from the kinematic region corresponding to edges of Dalitz plots, where the two light mesons move almost parallelly.
It is possible to catch the dominant contributions in a simple way through
introducing a new nonperturbative inputs, two-meson distribution amplitudes \cite{diehl,polyakov,mller,maul}. This just is one crucial step for the perturbative QCD approach (PQCD) to deal with the
three-body B meson decays, which are simplified into the two-body decays.
The two-meson DAs (distribution amplitudes) describe the hadronization of two collinear quarks, together with another two quarks
popped out of the vacuum, into two collimated mesons. For the $PP$ system, the two-meson DAs from not only S wave \cite{lihn0,lihn1,lihn2,yli,ajma,zrui,zrui1}
but also P wave \cite{hnli3,yli2,ajma2,yli3,ajma3,yli4,ajma4} have been studied and used widely in $B$ meson decays. For $VP$ system, the two-meson DAs with more complicated structures are studied by
few works \cite{hnli4,cwang}.

In view of the status about $B$ meson three body decays, we would like to study the three body radiative decay $B_s\to \phi \bar K^0 \gamma$ in pQCD approach. In this decay, we can avoid the
aforementioned difficulties
such as the entangled nonresonant and resonant contributions, significant final-state interactions. At the same time, it provides a clean platform to study the two-meson distribution amplitudes
for the $\phi K$ pair. Beacause, the similar decay $B\to \phi K \gamma$ has been researched by experiments\cite{belle0,belle4,babar0} and theories\cite{hnli4,cwang}. These studies shows that
no clear evidence is found for the existence of a kaonic resonance decaying to $\phi K$. In pQCD approach\cite{cwang}, the decay $B\to \phi K \gamma$ was studied by introducing the two-meson
DAs for the $\phi K$
pair, where some parameters are extracted from the data \cite{pdg16}. So it is meaningful to check if the $\phi K$ pair DAs can be used in the decay $B_s\to \phi K \gamma$
to predict its branching ratio and direct CP violation.

The layout of this paper is as follows, we analyze the decay $B_s\to \phi \bar K^0 \gamma$ using the perturbative QCD approach in Section II.
The numerical results and discussions are given in
Section III, where the theoretical uncertainties are also
considered. The conclusions are presented in the final part.
\section{the perturbative calculations}
The effective Hamiltonian which includes flavor changing $b\to
d\gamma$ transition is given by \cite{buchalla}
\be
H_{eff}=\frac{G_F}{\sqrt{2}}\left[\sum_{q=u,c}V_{qb}V^*_{qd}\left(C_1(\mu)O^q_1(\mu)+C_2(\mu)O_2^q(\mu)\right)\right.\non \left.-V_{tb}V^*_{td}\sum_{i=3\sim8g}C_i(\mu)O_i(\mu)\right]+h.c.
\en

 Under the light-cone coordinates, the momentums of $B$ meson
and the $\phi K$ pair can be written as: \be
P_{B_s}=\frac{M_{B_s}}{\sqrt{2}}(1,1,0_T),\;\;\;\;
P=\frac{M_{B_s}}{\sqrt2}(1,\eta,0_T), \en where the parameter
$\eta=\omega^2/M^2_{B_s}$ with $\omega$ being the invariant mass of the
$\phi K$ pair. If we define $P_1$ and $P_2$ as the momenta of the
$\phi$ and $K$ mesons, respectively, we have $P=P_1+P_2$, with \be
P^+_1&=&\zeta P^+, P^-_1=[(1-\zeta)\eta+r^2_\phi]P^+, P_{1T}=\sqrt{(\zeta\omega^2-m^2_\phi)(1-\zeta)},\\
P^+_2&=&(1-\zeta)P^+, P^-_2=(\zeta\eta-r^2_\phi)P^+, P_{2T}=-P_{1T}=-\sqrt{(\zeta\omega^2-m^2_\phi)(1-\zeta)}.
\en
where $\zeta$ is the $\phi$ meson momentum fraction and the mass ratio $r_\phi=m_\phi/m_{B_s}$.
The on-shell condition $P^2_1=m^2_\phi, P^2_2=0$ is used to obtained $P_{1T(2T)}$.
The momentums of the spectator quarks in $B_s$ and $K$ mesons can be defined as:
\be
k_1=(0,\frac{M_{B_s}}{\sqrt2}x_1,\textbf{k}_{1T}),\;\;
k_2=(\frac{M_{B_s}}{\sqrt2}z,0,\textbf{k}_{2T}).
\en
The momentum of the photon $\gamma$ in the final state is written as $P_3=(M_{B_s}/\sqrt2)(0,1-\eta,0_T)$.
The transverse polarization vectors of  $\gamma$ and $\phi K$ pair are given as:
\be
\epsilon^*_\gamma(\pm)&=&\left(0,0,\frac{1}{\sqrt2}(\mp1,-i)\right),\\
\epsilon^*_{\phi
K}(\pm)&=&\left(0,0,\frac{1}{\sqrt2}(\pm1,-i)\right). \en

  The $\phi K$ pair distribution amplitudes can be related to the $\phi$ and kaon distribution amplitudes \cite{pball,braun} through calculating perturbatively the matrix elements
$\langle\phi(P_1,\epsilon_\phi)K^+(P_2)|\bar
u(y^-)\Gamma s(0)|0\rangle$,
 where $\Gamma=\gamma_\mu\gamma_5, \sigma_{\mu\nu}\gamma_5, \gamma_5, \gamma_\mu, I$
represent the different Lorentz structures. The matrix elements can be written as
the products of the kinematic factors with the corresponding form
factors \cite{hnli4}:
\be
\langle\phi \bar K^0|
\bar s(y^-)\gamma_{\mu}\gamma_5
 d(0)|0\rangle&=&P_\mu \int^1_0dze^{izP\cdot y}
\phi_\parallel(z,\zeta,\omega)+\omega\epsilon^*_{T{\mu}}\int^1_0dz e^{izP\cdot y}\Phi_a(z,\zeta,\omega),\\
\langle\phi \bar K^0|\bar
s(y^-)\sigma_{\mu\nu}\gamma_5
d(0)|0\rangle&=&-i\left\{(\epsilon_{T\mu} P_{1\nu}-\epsilon_{T\nu}
P_{1\mu})\int^1_0dze^{izP\cdot y}
\phi_T(z,\zeta,\omega)\right.\non &&
\left.+\frac{2}{\omega}(P_{1\mu}P_{2\nu}-P_{1\nu}P_{2\mu})\int^1_0dz e^{izP\cdot y}\Phi_3(z,\zeta,\omega) \right\}\\
\langle\phi \bar K^0 |\bar s(y^-)\gamma_5 d(0)|0\rangle&=&\omega\int^1_0dz e^{izP\cdot y}\Phi_p(z,\zeta,\omega),\\
\langle\phi \bar K^0|\bar s(y^-)\gamma_\mu d(0)|0\rangle&=&i\frac{\omega}{P\cdot n_-}\epsilon_{\mu\nu\rho\sigma}\epsilon^\nu_T P^\rho n^\sigma_-\int^1_0dz e^{izP\cdot y\Phi_v(z,\zeta,\omega)},\\
\langle\phi \bar K^0 |\bar s(y^-)I d(0)|0\rangle&=&0.
\en
Some explanations are in order. The perturbative calculation is only in order to obtain the $\zeta$ dependence of the $\phi K$ pair distribution amplitudes, which arises from the Lorentz structure
of the associated hadronic matrix element, irrespective of whether the form factor is nonperturbative. To obtain the upper expansions, the following approximations to the kinematic factors have been used
\be
(P_1-P_2)_\mu &\approx & (2\zeta-1)P_\mu,\\
\epsilon_{T\mu}(\phi) P_{1\nu}-\epsilon_{T\nu}(\phi) P_{1\mu}&\approx & \zeta(\epsilon_{T\mu} P_{\nu}-\epsilon_{T\nu} P_{\mu}),\\
\frac{2}{\omega}(P_{1\mu}P_{2\nu}-P_{1\nu}P_{2\mu})&\approx &(2\zeta-1)(\epsilon_{L\mu} P_{\nu}-\epsilon_{L\nu} P_{\mu}),\\
\frac{2}{\omega}\epsilon_{\mu\nu\rho\sigma}\epsilon_T^\nu(\phi)P^\rho_1P^\sigma_2&\approx
& \frac{\omega}{P\cdot
n_{-}}(2\zeta-1)\sigma_{\mu\nu\rho\sigma}\epsilon^\nu_T P^\rho
n^\sigma_-. \en
Then the $\phi K$ pair distribution amplitudes up to
twist-3 can be given as \cite{hnli4}:
\be
\langle\phi K(P,
\epsilon_L)|\bar
u(y^-)_js(0)_l|0\rangle&=&\frac{1}{\sqrt{2N_c}}\int^1_0dze^{izP\cdot
y}\left\{(\gamma_5\psl)_{lj}\phi_\parallel(z,\zeta,\omega)+(\gamma_5)_{lj}\omega\phi_p(z,\zeta,\omega)\right.\non &&\left.+(\gamma_5\esl_L\psl)_{lj}\phi_3(z,\zeta,\omega)\right\},\\
\langle\phi K(P,
\epsilon_T)|\bar
u(y^-)_js(0)_l|0\rangle&=&\frac{1}{\sqrt{2N_c}}\int^1_0dze^{izP\cdot
y}\left\{(\gamma_5\esl_T\psl)_{lj}\phi_t(z,\zeta,\omega)+(\gamma_5\esl_{T\mu})_{lj}\omega\right.\non&&
\left.\times \phi_a(z,\zeta,\omega)+i\frac{\omega}{P\cdot
n_-}\epsilon_{\mu\nu\rho\sigma}(\gamma^\mu)_{lj}\epsilon^\nu_TP^\rho
n^\sigma_-\Phi_v(z,\zeta,\omega)\right\}. \en where the transversely polarized distribution amplitudes can be expressed as the products
of the time-like form facors $F(\omega)$ and the $z,\zeta$- dependent functions:
\be
\phi_t(z,\zeta,\omega)&=&\frac{3F_T(\omega)}{\sqrt{2N_c}}z(1-z)\zeta,\\
\phi_a(z,\zeta,\omega)&=&\frac{3F_a(\omega)}{\sqrt{2N_c}}z(1-z)\left[1+a_1C^{3/2}_1(2z-1)\right],\\
\phi_v(z,\zeta,\omega)&=&\frac{3F_v(\omega)}{\sqrt{2N_c}}z(1-z)(2\zeta-1).
\en
Here the $z$ dependence of each DAs except $\phi_a$ is assumed to be asymptotic form $z(1-z)$. In order to make the $\phi K$ pair DA $\phi_a$ a bit asymmetric, the first Gegenbauer moment
is included.  For our considered decays, only the transverse components are used. It is similar for the longitudinal polarized distribution amplitudes
\be
\phi_\parallel(z,\zeta,\omega)&=&\frac{3F_\parallel(\omega)}{\sqrt{2N_c}}z(1-z)(2\zeta-1),\\
\phi_p(z,\zeta,\omega)&=&\frac{3F_p(\omega)}{\sqrt{2N_c}}z(1-z),\\
\phi_3(z,\zeta,\omega)&=&\frac{3F_3(\omega)}{\sqrt{2N_c}}z(1-z)(2\zeta-1).
\en
 Here the time-like form factors $F_{T,2,v}(\omega)$ are used to
define the normalization of different twist distribution amplitudes,
\be
F_{T,\parallel}(\omega)&=&\frac{m^2_{T\parallel}}{(\omega-m_l)^2+m^2_{T\parallel}},\\
F_{a,v,p,3}(\omega)&=&\frac{m_0m^2_{T\parallel}}{(\omega-m_l)^3+m_0m^2_{T\parallel}}, \label{fvv}
\en
where $m_0\approx1.7 GeV$ is the chiral scale and the threshold invariant mass $m_l=m_\phi+m_K$. The parameters $m_\parallel$ and $m_T$ are associated with the longitudinally and transversely polarized $\phi$ meson. Both of them are expected to be
few GeV \cite{lihn0} and can be determined
by fitting the measured branching ratios of $\tau \to \phi K \nu$ and $B \to \phi K \gamma$ \cite{pdg16}. In the later calculations, we set $m_T=m_\parallel=m_{T\parallel}=3$ GeV.
To construct these form factors, these two points must be considered:
First, respecting the kinematic threshold of the decay spectra, $F_{i}(m_l)=1, i=T,\parallel,a,v,p,3$; Second, the power behaviors of the form factors in the asymptotic region
with large $\omega$ satisfy $F_{T,\parallel}(\omega)\approx 1/\omega^2, F_{a,v,p,3}(\omega)\approx m_0/\omega^3$ \cite{keum,huang}.

For the wave function of the heavy $B$ meson, we take
\be
\Phi_{{B_s}}(x,b)= \frac{1}{\sqrt{2N_c}} (\psl_{{B_s}} +m_{{B_s}}) \gamma_5
\phi_{{B_s}} (x,b). \label{bmeson} \en Here only the contribution of
Lorentz structure $\phi_{{B_s}} (x,b)$ is taken into account, since the
contribution of the second Lorentz structure $\bar \phi_{{B_s}}$ is
numerically small and has been neglected. For the
distribution amplitude $\phi_{{B_s}}(x,b)$ in Eq.(\ref{bmeson}), we
adopt the following model: \be
\phi_{{B_s}}(x,b)=N_{{B_s}}x^2(1-x)^2\exp[-\frac{M^2_{{B_s}}x^2}{2\omega^2_b}-\frac{1}{2}(\omega_bb)^2],
\en where $\omega_b$ is a free parameter, we take
$\omega_b=0.5\pm0.05$ Gev in numerical calculations, and $N_{B_s}=63.7$
is the normalization factor for $\omega_b=0.5$.

The decay
spectrum is written as: \be
\frac{d\Gamma}{d\omega}=\tau_{B_s}\frac{G^2_F\omega|\vec{P}||\vec{P}_3|}{64\pi^3M^3_{B_s}}|\mathcal{M}|^2
=\tau_{B_s}\frac{G^2_F\sqrt{\eta}(1-\eta)|\mathcal{M}|^2}{256\pi^3}, \en
where the conditions $|\vec{P}||\vec{P_3}|=\frac{M^2_{B_s}}{4}(1-\eta),
\omega=\sqrt{\eta}M_{B_s}$ are used. For simplicity, we define the
common factor $F$ in each amplitude as \be
F_i=\frac{G_FeM^5_{B_s}C_F}{\sqrt2 \pi}V_i, \en where $C_F$ is color
factor, and $V_i$ as $V^*_{ib}V_{is}$. If both $\phi K$ pair and
$\gamma$ are left-handed, the explicit factorization formula from
$O_{7\gamma}$ operator can be written as: \be
\mathcal{M}^R_{7\gamma}&=&4F_i(1-\eta)\int_0^1 dx_1 dx_2\int_{0}^{\infty}b_1 d b_1 b_2 db_2\phi_{B_s}(x_1,b_1)\left\{
E_e(t_a)S_t(z)\right.\non &&\left.
\left[\sqrt{\eta}(2z-1)(\phi_a+\phi_v)-(z+1)\phi_t\right]h_e(x_1 z,z,b_1,b_2)\right.\non &&
\left.-\sqrt{\eta}(\phi_a+\phi_v)E_e(t_b)S_t(x_1)h_e(x_1 z,x_1-\eta,b_2,b_1)\right\},
\en
\begin{figure}[t]
\vspace{-4cm} \centerline{\hspace{2cm}\epsfxsize=18 cm
\epsffile{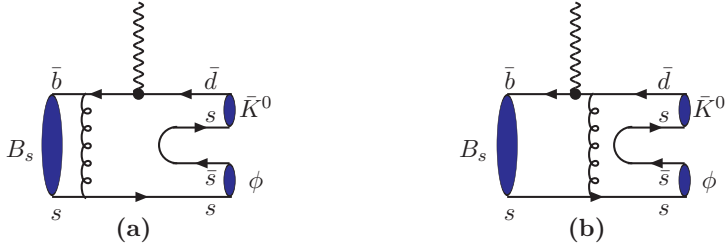}} \vspace{-18.8cm} \caption{ Feynman Diagrams of
electromagnetic penguin operator $O_{7\gamma}$. A photon is emitted
through the $O_{7\gamma}$ operator, and one hard gluon is exchanged
between an energy quark and the spectator quark.}
 \label{Figure1}
\end{figure}
where the hard scales are given as:
\be t_a&=&\max(\sqrt{z}M_B,\sqrt{\beta_{ab}}M_B,1/b_1,1/b_2),
t_b=\max(\sqrt{|\eta-x_1|}M_B,\sqrt{\beta_{ab}}M_B,1/b_1,1/b_2),\;\;\;\;\;\;  \en
and the hard function $h_e$ in the amplitude is from the propagators of
virtual quark and gluon: \be
h_e(\alpha,\beta,b_1,b_2)&=&\left[\theta(\beta)K_0(\sqrt{\beta}M_{B_s}b_1)+\theta(-\beta)i\frac{\pi}{2}H_0(\sqrt{-\beta}M_{B_s}b_1)\right],\non
&&
\times\left\{\theta(b_1-b_2)\left[\theta(\alpha)K_0(\sqrt{\alpha}M_{B_s}b_1)I_0(\sqrt{\alpha}M_{B_s}b_2)\right.\right.\non
&&
\left.\left.+\theta(-\alpha)\frac{i\pi}{2}H^{(1)}_0(\sqrt{-\alpha}M_{B_s}b_1)J_0(\sqrt{-\alpha}M_{B_s}b_2)\right]+(b_1\leftrightarrow
b_2)\right\}.\label{he1} \en
The Sudakov factor $S_t(x)$ from the threshold resummation is parameterized as
\be
S_t(x)=\frac{2^{1+2c}\Gamma(3/2+c)}{\sqrt{\pi}\Gamma(1+c)}[x(1-x)]^c,
\en
with $c=0.4$. The evolution factor is given as:
\be
E_e(t)=\alpha_s(t)\exp[-S_{B_s}(t)-S_{\phi K}(t)]
\en
with the Sudakov exponents $S_{B_s}(t), S_{\phi K}(t)$  being given in Ref.\cite{cwang}.
Compared with the left-helicity amplitude, the
right-helicity amplitude is proportional to the ratio $r_s=m_s/m_b$
which is obviously highly suppressed. $\mathcal{M}^L_{7\gamma}=0$ by
neglecting the mass of $s$ quark.

If the hard gluon which is required to kick the soft spectator quark
is generated from the $\mathcal{O}_{8g}$ operator, one can obtain
four diagrams from $O_{8g}$ operator given in Fig.3. The amplitudes
for the first two diagrams are listed as: \be
\mathcal{M}^{R,ab}_{8g}&=&2F_t\int_0^1 dx_1 dx_2\int_{0}^{\infty}b_1
d b_1 b_2 db_2\phi_{B_s}(x_1,b_1)\left\{
Q_bE_e(t_a^\prime)S_t(x_1)(\eta-x_1-1)\right.\non &&
\left.\left.\times[x_1\phi_t+\sqrt{\eta}z(\phi_a+\phi_v)\right]h_e(\alpha_a,\beta_{ab},b_2,b_1)
+Q_sE_e(t_b^\prime)S_t(z)\right.\non && \left.\times\left[(x_1-2z)\phi_t+3z\sqrt{\eta}(\phi_a+\phi_v)\right]h_e(\alpha_b,\beta_{ab},b_1,b_2)\right\},\\
\mathcal{M}^{R,cd}_{8g}&=&2F_tQ_u(1-\eta)\int_0^1 dx_1 dx_2\int_{0}^{\infty}b_1 d b_1 b_2 db_2\phi_{B_s}(x_1,b_1)\left\{
E_e(t_c)S_t(x_1)\right.\non && \left.\left[(\phi_a+\phi_v)\sqrt{\eta}(x_1-z)+x_1\phi_t\right]h_e(\alpha_c,\beta_{cd},b_2,b_1)
\right.\non && \left.-E_e(t_d)S_t(z)\left[3z\sqrt{\eta}\times(\phi_a+\phi_v)-(x_1+2\eta-2-z)\phi_t\right]\right.\non &&\left.
\times h_e(\alpha_d,\beta_{cd},b_1,b_2)\right\},\\
\mathcal{M}^{L,ab}_{8g}&=&2F_tQ_s\int_0^1 dx_1 dx_2\int_{0}^{\infty}b_1 d b_1 b_2 db_2\phi_{B_s}(x_1,b_1)\left\{
E_e(t_b^\prime)S_t(x_1)(1-z)\right.\non && \left.\times\left[\eta(2x_1-z)\phi_t-3x_1\sqrt{\eta}(\phi_a-\phi_v)\right]h_e(\alpha_b,\beta_{ab},b_1,b_2)\right\},\\
\mathcal{M}^{L,cd}_{8g}&=&2F_tQ_u\int_0^1 dx_1 dx_2\int_{0}^{\infty}b_1 d b_1 b_2 db_2\phi_{B_s}(x_1,b_1)\left\{
E_e(t_c)S_t(x_1)\right.\non && \left.\times(1-\eta)\left[\eta(x_1-z)\phi_t+x_1\sqrt{\eta}(\phi_a-\phi_v)\right]h_e(\alpha_c,\beta_{cd},b_2,b_1)
\right.\non && \left.+E_e(t_d)S_t(z)\left[\eta z(2 x_1+\eta-2z-1)\phi_t-3\sqrt{\eta}(1-\eta)(z-x_1)(\phi_a-\phi_v)\right]\right.\non &&\left.
\times h_e(\alpha_c,\beta_{cd},b_1,b_2)\right\},
\en where
\be
\alpha_a&=&1-\eta+x_1, \alpha_b=z-1,\beta_{ab}=x_1z, \label{t11}\\
\alpha_c&=&(1-\eta)x_1, \alpha_d=-z(1-\eta), \beta_{cd}=(x_1-z)(1-\eta),\\
t^\prime_a&=&\max(\sqrt{\alpha_a}M_{B_s},\sqrt{\beta_{ab}}M_{B_s},1/b_1,1/b_2),\\
t^\prime_b&=&\max(\sqrt{-\alpha_b}M_{B_s},\sqrt{\beta_{ab}}M_{B_s},1/b_1,1/b_2),\\
t_c&=&\max(\sqrt{\alpha_c}M_{B_s},\sqrt{|\beta_{cd}|}M_{B_s},1/b_1,1/b_2),\\
t_d&=&\max(\sqrt{-\alpha_d}M_{B_s},\sqrt{|\beta_{cd}|}M_{B_s},1/b_1,1/b_2).\label{t22}
\en
\begin{figure}[]
\vspace{-4cm} \centerline{\hspace{2cm}\epsfxsize=18 cm
\epsffile{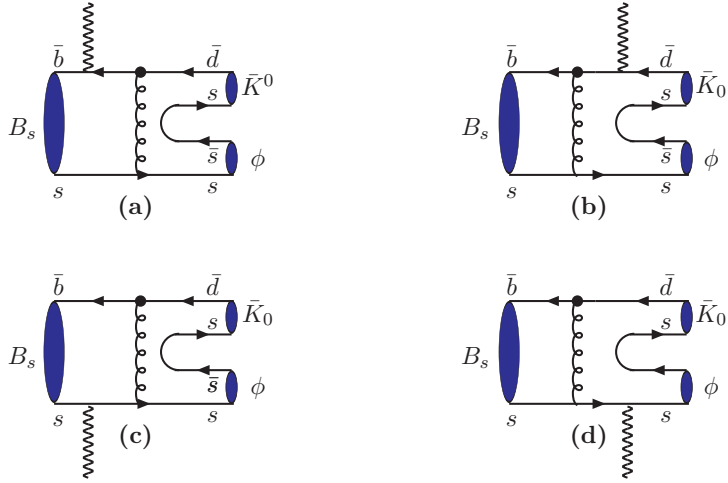}} \vspace{-15.8cm} \caption{ Feynman Diagrams of
chromomagnetic penguin operator $O_{8g}$. A hard gluon is emitted
through the $O_{8g}$ operator, and photon is emitted by bremsstrahlung of external quark lines.}
 \label{Figure1}
\end{figure}
There are two kinds of charm/up quark loop contributions specified as the type of photon emission: Quark line photon emission
and loop line photon emission. The former means that a photon is emitted through the external quark lines (shown in Fig.{\ref{fig3}}), the gauge invariant
$c/u$ quark loop function combined with the $ b\to  dg$ vertex is written as $\bar d \gamma^\mu(1-\gamma^5)I_{\mu\nu}b$ where
the explicit formula $I_{\mu\nu}$ is as follows \cite{bander}:
\be
I^a_{\mu\nu}&=&\frac{gT^a}{2\pi^2}(k^2g_{\mu\nu}-k_{\mu} k_{\nu})\int_0^1dx (1-x)\left[1+\log\frac{m^2_i-x(1-x)k^2}{t^2}\right]\non
&=&-\frac{gT^2}{8\pi^2}(k^2g_{\mu\nu}-k_\mu k_\nu)\left[G(m^2_i,k^2,t)-\frac{2}{3}\right],
\en
where $k$ is the gluon momentum and $m_i$ is the loop internal quark mass. The loop function $G$ is given as \cite{mat}:
\be
G(m^2_i,k^2,t)&=&\theta(-k^2)\frac{2}{3}\left[\frac{5}{3}+\frac{4m^2_i}{k^2}-\ln\frac{m^2_i}{t^2}+(1+\frac{2m^2_i}{k^2})\sqrt{1-\frac{4m^2_i}{k^2}}\ln\frac{\sqrt{1-4m^2_i/k^2}-1}{\sqrt{1-4m^2_i/k^2}+1}\right]
\non &&+\theta(k^2)\theta(4m^2_i-k^2)\frac{2}{3}\left[\frac{5}{3}+\frac{4m^2_i}{k^2}-\ln\frac{m^2_i}{t^2}-2(1+\frac{2m^2_i}{k^2})\right.\non &&\left.
\sqrt{\frac{4m^2_i}{k^2}-1}\arctan(\frac{1}{\sqrt{4m^2_i/k^2-1}})\right]
+\theta(k^2-4m^2_i)\frac{2}{3}\left[\frac{5}{3}+\frac{4m^2_i}{k^2}-\ln\frac{m^2_i}{t^2}\right.\non &&\left.+(1+\frac{2m^2_i}{k^2})\sqrt{1-\frac{4m^2_i}{k^2}}
\left[\ln\frac{1-\sqrt{1-4m^2_i/k^2}}{1+\sqrt{1-4m^2_i/k^2}}+i\pi\right]\right].
\en
It is noticed that there is no singularity when we take the limit of $k\to 0$, so we can neglect $k_T$ components of $k^2$ in the loop function $G$.
This kind contribution can be expressed as follows:
\be
\mathcal{M}^{R,ab}_{1i}&=&F_iQ_s\int_0^1 dx_1 dx_2\int_{0}^{\infty}b_1 d b_1 b_2 db_2\phi_{B_s}(x_1,b_1)\left[G(m^2_i,-\beta_{ab}M^2_{B_s},t^\prime_b)-\frac{2}{3}\right]\non &&
E_e(t_b^\prime)S_t(z)z\left[\sqrt{\eta}(2x_1-z)(\phi_a+\phi_v)-3x_1\phi_t\right]h_e(\alpha_b,\beta_{ab},b_1,b_2),\\
\mathcal{M}^{L,ab}_{1i}&=&F_i\int_0^1 dx_1 dx_2\int_{0}^{\infty}b_1 d b_1 b_2 db_2\phi_{B_s}(x_1,b_1)\sqrt{\eta}\left\{
Q_b\left[G(m^2_i,-\beta_{ab}M^2_{B_s},t^\prime_a)-\frac{2}{3}\right]\right.\non &&\left.
\times E_e(t^\prime_a)S_t(x_1)z(1+x_1-\eta)\left[x_1(\phi_a-\phi_v)+\sqrt{\eta}z\phi_t\right]h_e(\alpha_a,\beta_{ab},b_2,b_1)
\right.\non &&\left.+\left[G(m^2_i,-\beta_{ab}M^2_{B_s},t_b^\prime)-\frac{2}{3}\right]x_1(z-1)\left[(x_1-2z)(\phi_a-\phi_v)+3\sqrt{\eta}z\phi_t\right]
\right.\non &&\left.\times Q_sE_e(t_b^\prime)S_t(z)h_e(\alpha_b,\beta_{ab},b_1,b_2)\right\},
\en
\be
\mathcal{M}^{R,cd}_{1i}&=&F_iQ_u\int_0^1 dx_1 dx_2\int_{0}^{\infty}b_1 d b_1 b_2 db_2\phi_{B_s}(x_1,b_1)(\eta-1)\left\{\left[G(m^2_i,-\beta_{cd}M^2_{B_s},t_c)-\frac{2}{3}\right]\right.\non &&\left.
\times E_e(t_c)S_t(x_1)(\eta-1)\left[\sqrt{\eta}(z-x_1)(\phi_a+\phi_v)-x_1\phi_t\right]h_e(\alpha_c,\beta_{cd},b_1,b_2)\right.\non &&\left.+
G(m^2_i,-\beta_{cd}M^2_{B_s},t_d)E_e(t_d)S_t(z)\left[3(\eta-1)(z-x_1)\phi_t+(\phi_a+\phi_v)\right.\right.\non && \left.\left.
\times z\sqrt{\eta}(\eta+2x_1-1-2z)\right]h_e(\alpha_d,\beta_{cd},b_1,b_2)\right\},\\
\mathcal{M}^{L,cd}_{1i}&=&F_iQ_u\int_0^1 dx_1 dx_2\int_{0}^{\infty}b_1 d b_1 b_2 db_2\phi_{B_s}(x_1,b_1)(\eta-1)(x_1-z)\sqrt{\eta}\left\{E_e(t_c)S_t(x_1)\right.\non &&
\left.\times G(m^2_i,-\beta_{cd}M^2_{B_s},t_c)\left[\sqrt{\eta}(z-x_1)\phi_t-x_1(\phi_a-\phi_v)\right]h_e(\alpha_c,\beta_{cd},b_1,b_2)\right.\non &&\left.+
G(m^2_i,-\beta_{cd}M^2_{B_s},t_d)E_e(t_d)S_t(z)\left[3\sqrt{\eta}z\phi_t+(\phi_a-\phi_v)\right.\right.\non && \left.\left.
\times (2+z-2\eta-x_1)\right]h_e(\alpha_d,-\beta_{cd},b_1,b_2)\right\},
\en
where the hard function and the hard scales have been defined in Eq.(\ref{he1}) and Eq.({\ref{t11}}$\sim$\ref{t22}), respectively.
When the photon is emitted from the $c/u$ quark loop line (shown in Fig.4), the $ b\to  d g \gamma$ amplitude is expressed as \cite{liuj,simma,chang}
\begin{figure}[]
\vspace{-4cm} \centerline{\hspace{2cm}\epsfxsize=18 cm
\epsffile{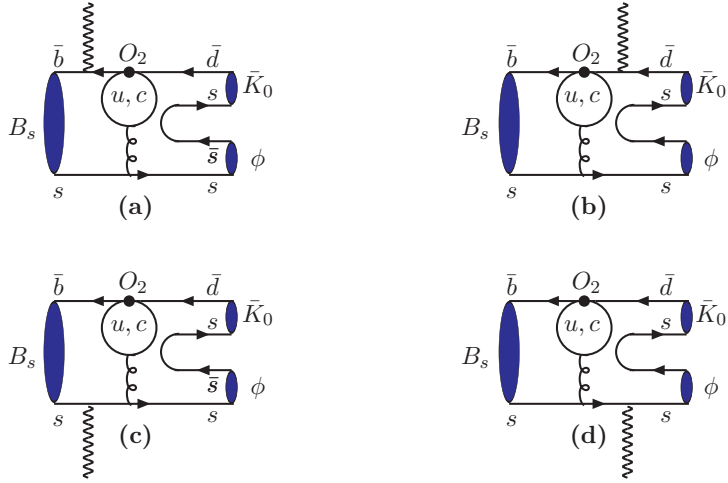}} \vspace{-15.8cm} \caption{ Feynman Diagrams of
quark line photon emission. The charm or up loop to gluon and attach to the spectator quark line. A photon is emitted through the external quark lines.}
 \label{fig3}
\end{figure}
\be
A( b\to d g\gamma)=\bar d \gamma^\rho\frac{(1-\gamma_5)}{2}T^abI_{\mu\nu\rho}\epsilon^\mu_\gamma\epsilon^\nu_{g},
\en
where the vertex function
$I_{\mu\nu\rho}$ is defined as follows,
\be
I_{\mu\nu\rho}&=&A_4\left[(k\cdot
q)\epsilon_{\mu\nu\rho\sigma}(q^\sigma-k^\sigma)
+\epsilon_{\nu\rho\sigma\tau}q^\sigma k^\tau k_\mu
-\epsilon_{\mu\rho\sigma\tau}q^\sigma
k^\tau q_\nu\right]\non &&
+A_5\left[\epsilon_{\mu\rho\sigma\tau}q^\sigma
k^\tau k_\nu
-k^2\epsilon_{\mu\nu\rho\sigma}q^\sigma\right],
\en \be
A_4&=&\frac{4i}{3\pi^2}eg\int_0^1 dx\int_0^{1-x} dy\frac{xy}{x(1-x)k^2+2xyq\cdot k-m^2_i},\\
A_5&=&\frac{4i}{3\pi^2}eg\int_0^1 dx\int_0^{1-x}
dy\frac{x(1-x)}{x(1-x)k^2+2xyq\cdot k-m^2_i}, \en
where $k$ is the gluon momentum and $q$ is the momentum of the photon. The amplitudes can be expressed as follows:
\be
\mathcal{M}^R_{2i}&=&\frac{8F_{i}}{3}\int_0^1 dx\int_0^{1-x} dy\int_0^1
dx_1dx_2\int_0^\infty b_1d b_1\phi_{B_s}(x_1,b_1)E_e(t_2)h^\prime_{e}(A,B,b_1)\non &&
\times\frac{(\eta-1)z}{xyz(1-\eta)M^2_{B_s}-m^2_i}\left\{xy\left[\sqrt{\eta}(\eta+z-x_1-1)(\phi_a+\phi_v)+(1+2x_1-\eta)\phi_t
\right]\right.\non &&\left.-x(1-x)\left[3x_1\phi_t+\sqrt{\eta}(z-2x_1)(\phi_a+\phi_v)\right]\right\},\\
\mathcal{M}^L_{2i}&=&\frac{8F_{i}}{3}\int_0^1 dx\int_0^{1-x} dy\int_0^1
dx_1dx_2\int_0^\infty b_1d b_1\phi_{B_s}(x_1,b_1)E_e(t_2)h^\prime_{e}(A,B,b_1)\non &&
\times\frac{(\eta-1)z}{xyz(1-\eta)M^2_{B_s}-m^2_i}\left\{xy\left[\sqrt{\eta}(z-x_1)(\phi_a-\phi_v)-2\eta z\phi_t
\right]\right.\non &&\left.+x(1-x)\left[\eta z\phi_t+\sqrt{\eta}x_1(\phi_a-\phi_v)\right]\right\},
\en where
\begin{figure}[]
\vspace{-4cm} \centerline{\hspace{2cm}\epsfxsize=18 cm
\epsffile{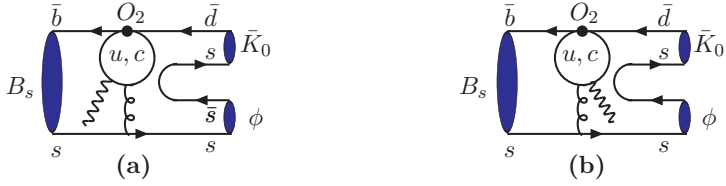}} \vspace{-18.8cm} \caption{ Feynman Diagrams of
loop line photon emission. The charm or up loop go to gluon and attach to the spectator quark line.
A photon is emitted through the quark loop line.}
 \label{fig4}
\end{figure}
\be
A&=&x_1 z, B=x_1z-\frac{y}{1-x}z(1-\eta)+\frac{r^2_i}{x(1-x)},\\
t_2&=&\max(\sqrt{A}M_{B_s},\sqrt{|B|}M_{B_s},1/b_1),\\
h^\prime_e(A,B,b_1)&=&K_0(\sqrt{A}M_{B_s}b_1)-\left[\theta(B)K_0(\sqrt{B}M_{B_s}b_1)
+\theta(-B)\frac{i\pi}{2}H_0(\sqrt{-B}M_{B_s}b_1)\right],
\en
where the ratio $r^2_i=m^2_i/M^2_{B_s}$ with $m_i$ being the $c/u$ quark mass.

For the annihilation diagrams, there are three kinds of operators are used: Both  $\bar b\otimes \bar s$ and $\bar d\otimes s$
are left-handed currents, denoted as $LL$; $\bar b \otimes \bar s$ is left-handed current and $\bar d \otimes s$ is right-handed
current, denoted as $LR$; the third kind of current $SP$ is from the Fierz transformation of $LR$ current.
So the factorization formulae for these annihilation diagrams are written as:
\begin{figure}[]
\vspace{-4cm} \centerline{\hspace{2cm}\epsfxsize=18 cm
\epsffile{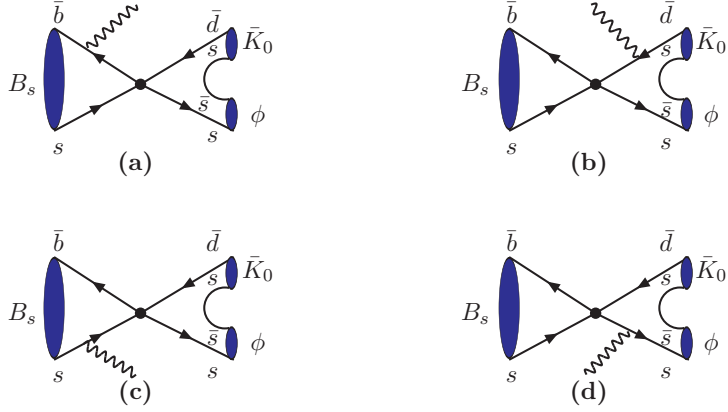}} \vspace{-15.8cm} \caption{ Annihilation  diagrams by tree and penguin operators.
There are no hard gluons for these cases because they belong the four Fermi interactions and do not include the spectator quarks.}
 \label{Figure1}
\end{figure}
\be
M^{R,LL}_{ann1}&=&4\sqrt{6\eta}eM^3_{B_s}Q_s(1-\eta)F_v\zeta\int_0^1
dx_1\int_{0}^{\infty}b_1 d b_1\phi_{B_s}(x_1,b_1)E_{a1}(t'_c)\non &&
\times S_t(x_1)K_0(\sqrt{\alpha_c}M_{B_s} b_1),\\
M^{L,LL}_{ann1}&=&4\sqrt{6\eta}eM^3_{B_s}F_v(1-\zeta)\int_0^1 dx_1\int_{0}^{\infty}b_1 d b_1\phi_{B_s}(x_1,b_1)S_t(x_1)
\left\{Q_b\right.\non &&\left.\times (\eta-x_1-1)E_{a1}(t_a{''})K_0(\sqrt{\alpha_a}M_{B_s}b_1)-Q_s x_1 E_{a1}(t'_c)K_0(\sqrt{\alpha_c}M_{B_s} b_1)\right\},
\en
\be
M^{R,LR}_{ann1}&=&4\sqrt{6\eta}eM^3_{B_s}Q_s(1-\eta)F_v(\zeta-1)\int_0^1
dx_1\int_{0}^{\infty}b_1 d b_1\phi_{B_s}(x_1,b_1)E_{a1}(t'_c)\non &&
\times S_t(x_1)K_0(\sqrt{\alpha_c}M_{B_s} b_1),\\
M^{L,LL}_{ann1}&=&-4\sqrt{6\eta}eM^3_{B_s}F_v\zeta\int_0^1 dx_1\int_{0}^{\infty}b_1 d b_1\phi_{B_s}(x_1,b_1)S_t(x_1)
\left\{Q_b\right.\non &&\left.\times (\eta-x_1-1)E_{a1}(t_a{''})K_0(\sqrt{\alpha_a}M_{B_s}b_1)-Q_s x_1 E_{a1}(t'_c)K_0(\sqrt{\alpha_c}M_{B_s} b_1)\right\},\\
M^{R,LL}_{ann2}&=&-M^{L,LR}_{ann2}=-2\sqrt{6\eta}ef_{B_s}M^3_{B_s}\int_0^1 d z\int_{0}^{\infty}b_2 d b_2 (\phi_a+\phi_v) S_t(z)\left\{
Q_dE_{a2}(t_b'')\right.\non &&\left.\times \frac{i\pi}{2}H^{(1)}_0(\sqrt{-\alpha_b}M_{B_s}b_2)-Q_s z\frac{i\pi}{2}E_{a2}(t_d')H^{(1)}_0(\sqrt{-\alpha_d}M_{B_s}b_2)\right\},\\
M^{L,LL}_{ann2}&=&-M^{R,LR}_{ann2}=2\sqrt{6\eta}ef_{B_s}M^3_{B_s}\int_0^1 d z\int_{0}^{\infty}b_2 d b_2 (\phi_a-\phi_v) \left\{
\frac{i\pi}{2}H^{(1)}_0(\sqrt{-\alpha_b}M_{B_s}b_2)\right.\non &&\left.\times (z-1)Q_dE_{a2}(t_b'')-Q_s (\eta-1)\frac{i\pi}{2}E_{a2}(t_d')H^{(1)}_0(\sqrt{-\alpha_d}M_{B_s}b_2)\right\}S_t(z),\\
M^{R,SP}_{ann2}&=&4\sqrt{6}ef_{B_s}M^3_{B_s}\int_0^1 d z\int_{0}^{\infty}b_2 d b_2 \phi_t \left\{ Q_dE_{a2}(t_b{''})
\frac{i\pi}{2}H^{(1)}_0(\sqrt{-\alpha_b}M_{B_s}b_2)\right.\non && \left.-Q_s (\eta-1)E_{a2}(t_d')\frac{i\pi}{2}H^{(1)}_0(\sqrt{-\alpha_d}M_{B_s}b_2)\right\}S_t(z),\\
M^{L,SP}_{ann2}&=&4\sqrt{6}ef_{B_s}M^3_{B_s}\int_0^1 d z\int_{0}^{\infty}b_2 d b_2 \phi_t \left\{ (1-z)\eta Q_dE_{a2}(t_b{''})
\frac{i\pi}{2}H^{(1)}_0(\sqrt{-\alpha_b}M_{B_s}b_2)\right.\non && \left.+Q_s \eta z E_{a2}(t_d')\frac{i\pi}{2}H^{(1)}_0(\sqrt{-\alpha_d}M_{B_s}b_2)\right\}S_t(z),
\en
where the time-like form factor $F_v$ is given in Eq.(\ref{fvv}), the hard scales $t_a'', t_b'', t_c', t_d'$ and the evolution factors $E_{ann1}, E_{ann2}$ are defined as:
\be
t_a''&=&\max(\sqrt{\alpha_a}M_{B_s},1/b_1),t_b''=\max(\sqrt{-\alpha_b}M_{B_s},1/b_2),\\
t_c'&=&\max(\sqrt{\alpha_c}M_{B_s},1/b_1),t_d'=\max(\sqrt{-\alpha_d}M_{B_s},1/b_2),\\
E_{ann1}(t)&=&\alpha_s(t)\exp[-S_{B_s}(t)],E_{ann2}(t)=\alpha_s(t)\exp[-S_{\phi K}(t)].
\en
By combining these amplitudes from the different Feynman diagrams, one can get the total decay amplitude for the
decay $B_s\to \phi \bar K^0\gamma$:
\be
M^i(B_s\to \phi \bar K^0\gamma)&=&V^*_{ub}V_{us}C_2\left(\mathcal{M}^{j,ab}_{1u}+\mathcal{M}^{j,cd}_{1u}+\mathcal{M}^{j}_{2u}\right)\non &&
+V^*_{cb}V_{cs}C_2\left(\mathcal{M}^{j,ab}_{1c}+\mathcal{M}^{j,cd}_{1c}+\mathcal{M}^{j}_{2c}\right)\non &&
-V^*_{tb}V_{ts}\left[C_{7\gamma}\mathcal{M}^j_{7\gamma}+C_{8g}(\mathcal{M}^{j,ab}_{8g}+\mathcal{M}^{j,cd}_{8g})\right.
\non &&\left.
+(a_4-\frac{1}{2}a_{10})\left(\mathcal{M}^{j,LL}_{ann1}+\mathcal{M}^{j,LL}_{ann2}\right)+(a_6-\frac{a_8}{2})\mathcal{M}^{j,SP}_{ann2}\right],
\en
where the combinations of the Wilson coefficients are defined as
\be
a_4&=&C_4+C_3/3, a_6=C_6+C_5/3,\\
a_8&=&C_8+C_7/3,a_{10}=C_{10}+C_9/3.
\en
\section{the numerical results and discussions}
\begin{figure}[]
\vspace{-1cm} \centerline{\epsfxsize=14 cm \epsffile{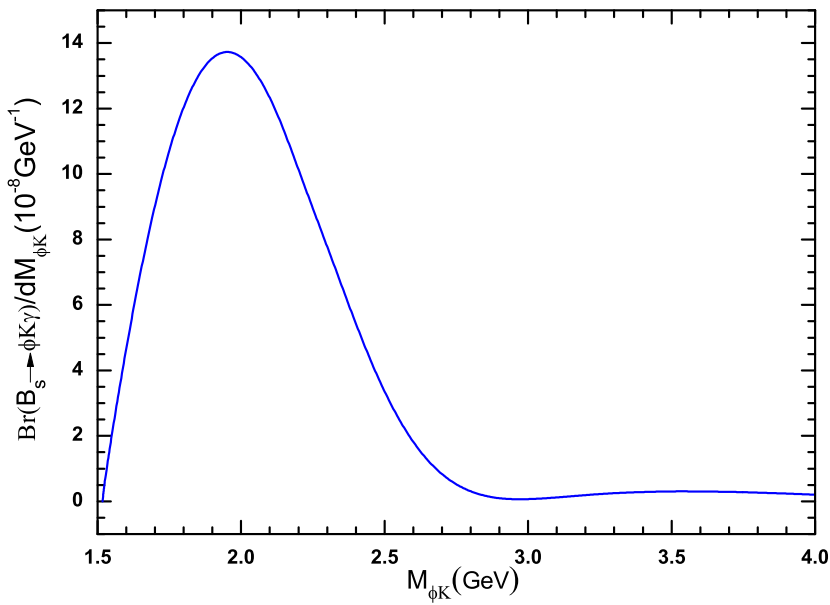}}
\vspace{-1.1cm} \caption{ $B_s\to \phi \bar K^0\gamma$ decay spectrum in the $\phi K$ invariant mass.}
 \label{fig6}
\end{figure}

The input parameters in the numerical calculations \cite{pdg16,hfag} are listed as following:
\be
f_{B_s}&=&230 MeV, \tau_{B_s}=1.510\times 10^{-12} s, M_{B_s}=5.3667 GeV\\
\lambda&=&0.22506\pm0.00050,  A=0.811\pm0.026, \bar\rho=0.124^{+0.019}_{-0.018}, \\
\bar\eta&=&0.356\pm0.011. \en Using the input parameters and the
wave functions as specified in this section and Sec.\ref{intro}, it
is easy to get the branching ratios for the decays $B_s\to \phi \bar K^0
\gamma$
\be Br(B_s\to \phi \bar K^0\gamma)&=&(9.26^{+1.79+3.12+0.64}_{-1.61-3.86-0.49})\times10^{-8}, \label{bran}
\en where the first error is from the $B_s$ meson wave function shape
parameter $\omega_b=(0.5\pm0.05)$ GeV, the second one is from the hard scale $t$ from $0.5t$ to $2t$
(without changing $1/b_i$), and the third one is from the Wolfenstein parameter $A=0.811\pm0.026$.
It is noticed that the result for the region with the $\phi K$ invariant mass as large as $4$ GeV is not
reliable, and the main contribution to the branching ratio is from the peak near the threshold value for
the $\phi K$ invariant mass as shown in Fig.{\ref{fig6}}. So the cut with $M_{\phi K}<2.5$ GeV is given to
the $\phi K$ invariant mass. From the result, one can find that the main error are from the scale-dependent uncertainty
, which can be reduced only if the next-to-leading order contributions are included. In the future, even if the high order corrections
are considered, the corresponding error is still large, it just reflects considerable nonperturbative effects. Certainly, we chose a wider range for the hard scale
from $0.5t$ to $2t$. It is noticed that the hard scale usually vary from $0.75t$ to $1.25t$ in pervious works.

We also predicted the $B_s\to \phi \bar K^0\gamma$ decay spectrum shown in Fig.6, which exhibits a maximum at the
$\phi K$ invariant mass around $1.95$ GeV.

The direct CP asymmetry of the $B_s\to \phi \bar K^0 \gamma$ is defined by
\be
A^{dir}_{CP}(B_s\to \phi K \gamma)=\frac{Br(\bar B^0_s\to \phi K^0 \gamma)-Br(B^0_s\to \phi \bar K^0 \gamma)}
{Br(\bar B^0_s\to \phi K^0 \gamma)+Br(B^0_s\to \phi \bar K^0 \gamma)}
\en
We predict the direct CP asymmetry as:
\be
A^{dir}_{CP}(B_s\to \phi \bar K^0 \gamma)=(-4.1^{+0.4+1.7+0.2}_{-0.6-1.2-0.1})\%,
\en
where the errors are the same with those in Eq.(\ref{bran}). The main error is still from the hard scale $t$ varying from $0.5t$
to $2t$. The errors induced by the hadronic uncertainties, such as the decay constants, the Gegenbauer moments and the shape parameter $\omega_b$, which
can be dropped out partly in the ratio. Similar conditions also occur in the direct CP asymmetries of the $B_s$ two body decays \cite{ali}.
Some comments are in order:
\begin{itemize}
\item
The branching ratio for the decay $B_s\to \phi \bar K^0\gamma$ is near $10^{-7}$ order, which can be tested by the present running
LHCb experiments. The ratio of the branching ratios for the decays $B\to \phi K$ and $B\to \phi K\gamma$ is given as \cite{pdg16}
\be
\frac{Br(B^+\to \phi K^+)}{Br(B^+\to \phi K^+\gamma)}&=&\frac{(8.8^{+0.7}_{-0.6})\times10^{-6}}{(2.7\pm0.4)\times10^{-6}}\sim3.3,\\
\frac{Br(B^0\to \phi K^0)}{Br(B^0\to \phi K^0\gamma)}&=&\frac{(7.3\pm0.7)\times10^{-6}}{(2.7\pm0.7)\times10^{-6}}\sim2.7,\\
\en
For the decay $B_s\to \phi \bar K^0$, it hasn't been measured by experiment, while studied by QCDF approach \cite{beneke} and PQCD approach \cite{ali}
and predicted as $(2.7^{+7.4}_{-2.5})\times10^{-7}$ and $(1.6^{+1.0}_{-0.5})\times10^{-7}$, respectively. Combined with our prediction we find that the ratio
\be
\frac{Br(B_s\to \phi \bar K^0)}{Br(B_s\to \phi \bar K^0\gamma)}\sim2.9 \;\text{or} \;1.7,
\en
that is to say, the branching ratios of decays $B_{u,d,s}\to \phi K$ are larger than that of corresponding  one photon radiation decays about $1\sim2$ times.
\item
For the $B_s\to \phi \bar K^0\gamma$ decay, the nonresonant contributions are dominant, and the resonant contributions from
$K_1(1650), K_2(1770)$ and $K(1830)$ mesons through the $\phi K$ channel are expected to be negligible. Because the branching ratios of
these resonant mesons decaying into $\phi K$ pair are not yet available. So this decay $B_s\to \phi \bar K^0\gamma$ provides a clean test
for the application of two-meson distribution amplitudes to the B meson three-body  decays. It is similar to the decay $B\to \phi K\gamma$.
\item
For the decay $B_s\to \phi \bar K^0\gamma$, the prominent feature of the decay spectra is the
enhancement near the threshold, which reaches the maximum at around $m_{\phi K}\sim1.95 GeV$. Compared the decay spectra of the
$B^0\to \phi K^0\gamma$ channel, the peak position move toward the larger $m_{\phi K}$ region because $B_s$ is heavier than $B$ meson.
Indeed, the decay $B^0\to \phi K^0\gamma$ spectrum exhibits a maximum at the $\phi K$ invariant mass around $1.8 GeV$ \cite{cwang}. However,
the shapes of the differential decay rates  for the two decay channels should be very similar.
\item
In the $b\to d\gamma$ process, although the dominant contribution to decay amplitudes comes from the chiral-odd dipole operator $O_{7\gamma}$,
we also considered contributions from $O_{8g}$ operator, the annihilation type amplitudes, especially $O_2$ operator from the quark loop corrections,
which is necessary to induce the direct CP violation. The CKM matrix element for the $O_{7\gamma}$ is $V^*_{tb}V_{td}\sim \lambda^3$, while the tree
operator $O_2$ is either proportional to $V^*_{ub}V_{ud}\sim\lambda^3$ or $V^*_{cb}V_{cd}\sim\lambda^3$. Then the tree contribution is not suppressed
and can give bigger direct CP asymmetries compared that of the decay $B^0\to \phi K^0\gamma$. For the $B^0\to \phi K^0\gamma$ decay, the CKM matrix for the $O_{7\gamma}$
and tree operators are proportional to $V^*_{tb}V_{ts}\sim\lambda^2$ and $V^*_{ub}V_{ud}\sim\lambda^4$, respectively. The difference between these two interfering
amplitudes is large, so the corresponding direct CP asymmetry is small \cite{cwang}.
\item
U-spin can connect $b\to s\gamma$ and $b\to d\gamma$ these two different kinds of weak decays by exchange of $d\leftrightarrow s$ \cite{soares,gronau,hurth}. Using the U-spin symmetry and CKM
unitarity relation
\be
Im(V^*_{ub}V_{us}V_{cd}V^*_{cs})=-Im(V^*_{ub}V_{ud}V_{cd}V^*_{cd}),
\en
one can obtain
\be
|\mathcal{A}(\bar B^0\to \phi \bar K^0\gamma)|^2-|\mathcal{A}( B^0\to \phi K^0\gamma)|^2=|\mathcal{A}( B_s\to \phi \bar K^0\gamma)|^2-|\mathcal{A}( \bar B_s\to \phi K^0\gamma)|^2.\;\;\;
\en
Certainly, U-spin symmetry breaking is introduced through the form factors $F_{B\to \phi K}$ and $F_{B_s\to \phi K}$. So we can expect that
\be
A^{dir}_{CP}(B_s\to \phi \bar K^0\gamma)\approx A^{dir}_{CP}(B^0\to \phi K^0\gamma)\frac{Br(B^0\to \phi K^0\gamma)}{Br(B_s\to \phi \bar K^0\gamma)}.
\en
Combined the predictions for the decay $B^0\to \phi K^0\gamma$ given in Ref.\cite{cwang}, one can find this relation is well supported.

\end{itemize}

\section{Conclusion}\label{summary}

In this paper, we calculate the branching ratio and the direct CP asymmtry for the decay $B_s\to \phi \bar K^0\gamma$, which induced by $b\to d \gamma$ transition. In addition to the dominant eletromagnetic penguin operator, the subleading contributions including the chromomagnetic penguin operator, quark-loop corrections and annihilation
amplitudes are also calculated.
Compared with the decay $B^0\to \phi K^0\gamma$, the branching ratio for the decay $B_s\to \phi \bar K^0\gamma$ is much smaller and less than $10^{-7}$ because of the smaller CKM element matrix being proportional to
$\lambda^3$. Although the subleading contribution from the the quark loop corrections is small, it is necessary for the direct CP asymmetry, which is predicted as $(-4.1^{+0.4+1.7+0.2}_{-0.6-1.2-0.1})\%$.
These predictions can be well explained by using the U-spin asymmetry approach when combined the results for the decay $B^0\to \phi K^0\gamma$. We also give the shape of the differential decay rate, which
is similar with that for the decay $B^0\to \phi K^0\gamma$ but with different peak position for the $m_{\phi K}$. In the upper calculations, the two-meson $\phi K$ DAs are introduced to absorb the
infrared dynamics in the $\phi K$ pair, so the three-body decay amplitude can be factorized, just similar to the two-body case, into the convolution of the wave functions and hard kernels.
\section*{Acknowledgment}
This work is partly supported by the National Natural Science
Foundation of China under Grant No. 11347030, the Program of
Science and Technology Innovation Talents in Universities of Henan
Province 14HASTIT037, and the Research Foundation of the Young Core Teacher from Henan University of Technology. Z.Q. Zhang is grateful to Prof. Hsiang-nan Li for helpful discussions.


\begin{thebibliography}{99}
\bibitem{belle0}
A. Drutskoy {\it et al.} (Belle Collaboration), \prl{\bf 92}, 051801 (2004).
\bibitem{belle1}
A. Garmash {\it et al.} (Belle Collaboration), \prd{\bf71}, 092003 (2005).
\bibitem{belle2}
A. Garmash {\it et al.} (Belle Collaboration), \prl{\bf96}, 251803 (2006).
\bibitem{belle3}
A. Garmash {\it et al.} (Belle Collaboration), \prd{\bf75}, 012006 (2007).
\bibitem{belle4}
H. Sahoo {\it et al.} (Belle Collaboration), \prd{\bf84}, 071101(R) (2011).
\bibitem{babar0}
B. Aubert {\it et al.} (BABAR Collaboration), \prd{\bf75}, 052005 (2007).
\bibitem{babar1}
B. Aubert {\it et al.} (BABAR Collaboration), \prd{\bf78}, 052005 (2008).
\bibitem{babar2}
B. Aubert {\it et al.} (BABAR Collaboration), \prd{\bf79}, 072006 (2009).
\bibitem{babar3}
P. Del Amo Sanchez {\it et al.} (BABAR Collaboration), \prd{\bf82}, 031101 (2010).
\bibitem{lhc1}
R. Aaij, {\it et al.} (LHCb Collaboration), \prl{\bf111}, 101801 (2013).
\bibitem{lhc2}
R. Aaij, {\it et al.} (LHCb Collaboration), \prd{\bf87}, 112009 (2013).
\bibitem{lhc3}
R. Aaij, {\it et al.} (LHCb Collaboration), \prl{\bf112}, 011801 (2014).
\bibitem{gronau1}
M. Gronau and J. L. Rosner, \plb{\bf564}, 90 (2003); \prd{\bf72}, 094031 (2005).
\bibitem{gronau2}
M. Gronau, \plb{\bf727}, 136 (2013).
\bibitem{dxu1}
D. Xu, G. N. Li and X. G. He, \plb{\bf728}, 579 (2014)
\bibitem{xghe}
X. G. He, G. N. Li and D. Xu, \prd{\bf91}, 014029 (2015).
\bibitem{imb}
M. Imbeault and D. London, \prd{\bf84}, 056002 (2011).
\bibitem{wshou1}
C. K. Chua, W. S. Hou, S. Y. Tsai, \prd{\bf66}, 054004 (2002).
\bibitem{wshou2}
C.K. Chua, W. S. Hou, S. Y. Shiau,S. Y. Tsai, \prd{\bf67}, 034012 (2003).
\bibitem{wshou3}
C. K. Chua, W. S. Hou, \epjc{\bf27}, 555 (2003).
\bibitem{furman}
A. Furman, R. Kami$\acute{\text{n}}$ski, L. Lesniak, and B. Loiseau, \plb{\bf622}, 207 (2005).
\bibitem{eibennich}
B. EU-Bennich {\it et al.}, \prd{\bf74}, 114009 (2006).
\bibitem{leitner}
O. Leitner, J.-P. Dedonder, B. Loiseau, and R. Kami$\acute{n}$ski, \prd{\bf81}, 094003 (2010).
\bibitem{cheng1}
H. Y. Cheng, K. C. Yang, \prd{\bf66}, 054015 (2002).
\bibitem{cheng2}
H. Y. Cheng, C. K. Chua, and A. Soni, \prd{\bf76}, 094006 (2007).
\bibitem{cheng3}
H. Y. Cheng, C. K. Chua, \prd{\bf88}, 114014 (2013); {\bf89}, 074025 (2014).
\bibitem{yingli}
Y. Li, \prd{\bf89}, 094007 (2014).
\bibitem{cheng4}
H. Y. Cheng, C. K. Chua, Z. Q. Zhang, \prd{\bf94}, 094015 (2016).
\bibitem{lihn0}
C. H. Chen and H. n. Li, \plb{\bf561}, 258 (2003).
\bibitem{lihn1}
W. F. Wang, H. C. Hu, H. n. Li, and C. D. Lu, \prd{\bf89}, 074031 (2014).
\bibitem{lihn2}
W. F. Wang, H. n. Li, W. Wang and C. D. Lu, \prd{\bf91}, 094024 (2015).
\bibitem{yli}
A. J. Ma, Y. Li, W. F. Wang, Z. J. Xiao, \epjc{\bf76}, 675 (2016).
\bibitem{ajma}
A. J. Ma, Y. Li, W. F. Wang, Z. J. Xiao, Chin. Phys. C {\bf41}, 083105 (2017).
\bibitem{zrui}
Z. Rui, Y. Li, W. F. Wang, \epjc{\bf77}, 199 (2017).
\bibitem{zrui1}
Z. Rui, W. F. Wang, \prd{\bf97}, 033006 (2018).
\bibitem{hnli3}
W. F. Wang, H. n. Li, \plb{\bf763}, 29 (2016).
\bibitem{yli2}
Y. Li, A. J. Ma, Z. J. Xiao, W. F. Wang, \prd{\bf95}, 056008 (2017).
\bibitem{ajma2}
A. J. Ma, Y. Li, W. F. Wang, Z. J. Xiao, \npb{\bf923}, 54 (2017).
\bibitem{yli3}
Y. Li, A. J. Ma, Z. Rui, Z. J. Xiao, \npb{\bf924}, 745 (2017).
\bibitem{ajma3}
A. J. Ma, Y. Li, Z. J. Xiao, \npb{\bf926}, 584 (2018).
\bibitem{yli4}
Y. Li, A. J. Ma, W. F. Wang, Z. J. Xiao, \prd{\bf96}, 036014 (2017).
\bibitem{ajma4}
A. J. Ma, Y. Li, W. F. Wang, Z. J. Xiao, \prd{\bf96}, 093011 (2017).
\bibitem{hnli4}
C. H. Chen, H. n. Li, \prd{\bf70}, 054006 (2004).
\bibitem{cwang}
C. Wang, J. B. Liu, H. n. Li, C. D. Lu, \prd{\bf97}, 034033 (2018).
\bibitem{krankl}
S. Kr$\ddot{\text{a}}$nkl, T. Mannel, J. Virto, \npb {\bf899}, 247 (2015).
\bibitem{diehl}
M. Diehl, T. Gousset, B. Pire and O. Teryaev, \prl{\bf81}, 1782 (1998).
\bibitem{polyakov}
M. V. Polyakov, \npb{\bf555},231 (1999).
\bibitem{mller}
D. M$\ddot{\text{u}}$ller, D. Robaschik, B. Geyer, F.-M. Dittes, J. Horejsi, Fortsch. Phys. {\bf42}, 101 (1994).
\bibitem{maul}
M. Maul, Eur. Phys. J. C{\bf21}, 115 (2001).
\bibitem{pdg16}
Particle Data Group Collaboration, K. A. Olive {\it et al.}, Chin. Phys. C {\bf40}, 100001 (2016).
\bibitem{buchalla}
G. Buchalla, A. J. Buras and M. E. Lautenbacher, Rev. Mod. Phys. {\bf68}, 1125 (1996).
\bibitem{pball}
P. Ball, V. M. Braun, Y. Koike, and K. Tanaka,  \npb {\bf529}, 323 (1998).
\bibitem{braun}
V. M. Braun and I. E. Filyanov, Z. Phys. C{\bf48}, 239 (1990); P. Ball, JHEP {\bf9901}, 010 (1999).
\bibitem{keum}
Y. Y. Keum, H. n. Li, and A. I. Sanda, \plb{\bf504}, 6 (2001); \prd{\bf 63}, 054008 (2001); Y. Y. Keum and H. n. Li, \prd{\bf63}, 074006 (2001).
\bibitem{huang}
T. Huang, X. H. Wu, and M. Z. Zhou, \prd{\bf70}, 014013 (2004).
\bibitem{bander}
M. Bander, D. Silerman and A. Soni, \prl{\bf43}, 242 (1979).
\bibitem{mat}
M. Matsumori, A. I. Sanda and Y. Y. Keum, \prd{\bf72}, 014013 (2005).
\bibitem{liuj}
J. Liu and Y. P. Yao, \prd{\bf42}, 1485 (1990).
\bibitem{simma}
H. Simma and D. Wyler, \npb {\bf344}, 283 (1990).
\bibitem{chang}
C. H. Chang and H. n. Li, Phys.Rev.D {\bf55}, 5577(1997);M. Nagashima and H. n. Li, Phys. Rev. D {\bf56}, 034001 (2003).
\bibitem{hfag}
The online update at http://www.slac.stanford.edu/xorg/hfag.
\bibitem{beneke}
M. Beneke and M. Neubert, \npb {\bf675}, 333 (2003).
\bibitem{ali}
A. Ali {et al.}, Phys. Rev. D {\bf76}, 074018 (2007).
\bibitem{soares}
J. M. Soares, \npb{\bf367}, 575 (1991).
\bibitem{gronau}
M. Gronau, \plb{\bf492}, 297 (2000).
\bibitem{hurth}
T. Hurth and T. Mannel, \plb{\bf511}, 196 (2001).
\end{thebibliography}
\end{document}